\documentclass[%
reprint,
twocolumn,
superscriptaddress,
showpacs,preprintnumbers,
nofootinbib,
aps,
prd
]{revtex4}

\usepackage{amsmath}
\usepackage{amssymb}
\usepackage{mathtools}
\usepackage{hyperref}
\usepackage{graphicx}
\usepackage[dvips]{color}
\usepackage{ulem}
\usepackage{comment}
\usepackage{tikz}
\usepackage[caption=false]{subfig}

\graphicspath{
{./}
{./figures/}
{./figures/squarebubbles/}
{./figures/roundbubbles/}
{./figures/unscaled/}
{./figures/slices/}
{./figures/hubercomp-unified/}
{./figures/comp-scaled-fluid/}
{./figures/comp-scaled-scalar/}
{./figures/ubars/}
{./figures/scaling/}
}

\begin{document}

\title{Revisiting the envelope approximation: gravitational waves from
  bubble collisions}

\author{David J. Weir}
\email{david.weir@uis.no}
\affiliation{
Institute of Mathematics and Natural Sciences,
University of Stavanger,
4036 Stavanger,
Norway
}

\date{June 15, 2016}

\begin{abstract}
We study the envelope approximation and its applicability to
first-order phase transitions in the early universe. We demonstrate
that the power laws seen in previous studies exist independently of the
nucleation rate. We also compare the envelope approximation prediction
to results from large-scale phase transition simulations. For phase
transitions where the contribution to gravitational waves from scalar
fields dominates over that from the coupled plasma of light particles,
the envelope approximation is in agreement, giving a power spectrum of
the same form and order of magnitude. In all other cases the form and
amplitude of the gravitational wave power spectrum is markedly
different and new techniques are required.
\end{abstract}

\pacs{64.60.Q-, 04.30.-w, 03.50.-z, 95.30.Lz}

\maketitle

\section{Introduction}

With the upgrades of the LIGO and VIRGO gravitational wave
observatories~\cite{Accadia:2009zz,Harry:2010zz}, it was only a
  matter of time before an astrophysical source of gravitational waves
  was detected~\cite{Abbott:2016blz}. Cosmological sources of
gravitational waves also exist, and their detection would offer an
exciting new tool to study the physics of the early
universe. Proposals for several space-based gravitational wave
detectors are under development, with sensitivities sufficient to
detect cosmological sources of gravitational waves. In particular,
eLISA is scheduled for launch in 2034 and offers a realistic prospect
of detecting gravitational waves from cosmological
sources~\cite{Seoane:2013qna}, including first-order phase
transitions~\cite{Caprini:2015zlo}.

There has long been interest in the expected gravitational wave power
spectrum produced by a first-order phase transition at, for example,
the electroweak
scale~\cite{Kamionkowski:1993fg,Apreda:2001us,Grojean:2006bp,Huber:2008hg,Ashoorioon:2009nf,Kozaczuk:2014kva,Dorsch:2014qpa,Kakizaki:2015wua,Leitao:2015fmj}. There
is also a growing body of more recent work aimed at making predictions
for other scenarios where a first-order phase transition may be
detectable~\cite{Schwaller:2015tja,Jaeckel:2016jlh,Dev:2016feu}.

Early studies modelled the process as the collision of thin shells of
stress-energy, initially for vacuum
transitions~\cite{Kosowsky:1991ua,Kosowsky:1992vn} and then later for
thermal transitions~\cite{Kamionkowski:1993fg}. Further refinements in
Ref.~\cite{Huber:2008hg} have set the state of the art and produced a
robust form of the power spectrum that has been widely adopted in the
literature when making predictions. For thermal phase transitions,
there is good understanding of how much energy ends up in the plasma
of light particles around the bubble wall~\cite{Espinosa:2010hh}. The
set of simplifying assumptions going into these thin-shell
calculations is usually termed the envelope approximation. While the
power spectrum must be computed numerically in the envelope
approximation, there is some analytical understanding of the process
as well~\cite{Caprini:2007xq,Caprini:2009fx,Jinno:2016vai}.

Meanwhile, progress has also been made in understanding other
mechanisms giving rise to gravitational waves after first-order phase
transitions. These include
turbulence~\cite{Kosowsky:2001xp,Kahniashvili:2008pf,Caprini:2009yp}
and acoustic waves in the plasma of light
  particles~\cite{Hindmarsh:2013xza,Giblin:2014qia,Hindmarsh:2015qta}. In
particular, the acoustic wave source produces a dramatically different
power spectrum form with potentially much greater amplitude than
predicted by the envelope approximation. As of yet, no analytic
calculation can reproduce the results of
Refs.~\cite{Hindmarsh:2013xza,Giblin:2014qia,Hindmarsh:2015qta}, in
which the acoustic behaviour was explored primarily through
simulations of a coupled field-fluid model.

In Ref.~\cite{Huber:2008hg}, the power spectrum from the envelope
approximation was parametrised as
\begin{equation}
\label{eq:huberfit}
\Omega^{\text{env}}_{\text{GW}}(\omega) = \tilde{\Omega}^{\text{env}}_{\text{GW}} \frac{(a+b)
  \tilde{\omega}^b \omega^a}{b\tilde{\omega}^{(a+b)} + a\omega^{(a+b)}},
\end{equation}
with peak frequency $\tilde{\omega}$, peak
amplitude $\tilde{\Omega}^{\text{env}}_{\text{GW}}$, and two power-law exponents $a \in
[2.66,2.82]$ and $b \in [0.90, 1.19]$. The fraction of energy in
gravitational waves was found to be
\begin{equation}
\label{eq:peakfreq}
\tilde{\Omega}_\text{GW}^\text{env} \simeq \frac{0.11 v_\mathrm{w}^3}{0.42+v_\mathrm{w}^2} \left( \frac{H_*}{\beta}\right)^2 \frac{\kappa^2 \alpha_T^2}{(\alpha_T+1)^2}.
\end{equation}
Here, $v_\mathrm{w}$ is the wall velocity, $\kappa$ is the efficiency
factor gauging what fraction of vacuum energy ends up contributing to
the stress-energy localised at the bubble wall, and $\alpha_T$ is the
ratio of latent heat to radiation. The ratio $H_*/\beta$ determines
how quickly the transition proceeds: $\beta$ is the nucleation rate
and $H_*$ is the Hubble rate at the time of the transition.

\begin{figure}[tb]
\begin{centering}
\includegraphics[width=0.45\textwidth,clip=true]{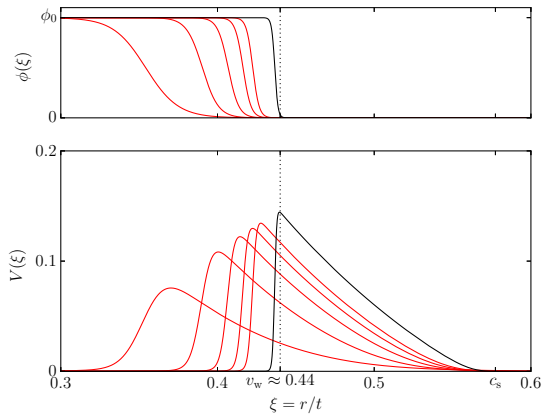}
\caption{\label{fig:scalingplot} Plot comparing (top) the radial field
  $\phi(\xi)$ and (bottom) fluid velocity $V(\xi)$ profiles for the
  `weak' parameters, as a function of $\xi = r/t$ (the parameters can
  be found in Table~\ref{tab:parameters}). A weak deflagration with
  $v_\mathrm{w} \approx 0.44$ is shown. The development of the
  profiles is illustrated by the red curves (at time intervals of
  $500/T_\mathrm{c}$ up to $2500/T_\mathrm{c}$), while the profile at
  $5000/T_\mathrm{c}$ is shown in black. The scalar field bubble wall
  remains of constant width, while the fluid profile approaches a
  scaling solution, and is of thickness $\sim \left|v_\mathrm{w} -
  c_\mathrm{s}\right| R_*/v_\mathrm{w}$ when bubbles of radius $R_*$
  collide. The envelope approximation of thin colliding shells might
  therefore be expected to work for scalar field walls, but plainly
  cannot for colliding fluid shocks unless $v_\mathrm{w} \approx
  c_\mathrm{s}$.}
\end{centering}
\end{figure}

In the calculations leading to the above expressions it was assumed
that the fluid kinetic energy was localised near the bubble wall,
whereas in reality it reaches a scaling profile proportional to the
bubble radius, meaning that the assumption of thin shells
instantaneously colliding does not hold (see
Fig.~\ref{fig:scalingplot}). Furthermore, the fluid kinetic energy
persists as sound waves in the plasma after the transition, until
turbulence and expansion attenuate this source.

In Ref.~\cite{Hindmarsh:2015qta}, it
was argued that the actual energy deposited in gravitational waves by
a fluid source is approximately
\begin{equation}
\label{eq:approxgwsound}
\tilde{\Omega}_\text{GW}^\text{sound} \simeq 3 \left( \kappa \alpha_T \right)^2 \left(H_*
(8\pi)^\frac{1}{3} \frac{v_\mathrm{w}}{\beta}\right) \tilde{\Omega}^{\text{env}}_{\text{GW}}
\end{equation}
-- typically a factor $60(\beta /H_*)$ larger than the envelope
approximation result. The power spectrum of gravitational waves was
also very different.

Therefore, the envelope approximation should not be used to fully
describe phase transitions where a lot of kinetic energy ends up in
the fluid. It may still remain valid for gravitational waves sourced
by the collision of scalar field bubble walls, which do not scale. It
may also model the initial collision of the fluid shells, if they are
thin enough, such as in Jouguet detonations.

The various types of sources -- scalar field collisions, fluid shell
collisions, acoustic waves and turbulence -- contribute to different
extents, depending on the model.

The principal aim of this paper is to directly test the envelope
approximation against a full lattice simulation for the first time,
concentrating on the scalar field bubble walls in a thermal phase
transition. This is motivated by two observations. First, collisions
of the scalar field walls will always source gravitational waves
(although the source may well be subdominant for a thermal phase
transition). Second, in certain cases, such as where the wall runs
away, scalar field collisions may be the dominant source of
gravitational waves.

We will also investigate the suitability of the envelope
approximation result for modelling the production of gravitational
waves by colliding plasma shells. Immediately after the shells have
collided, it does get the amplitude of gravitational waves
approximately right, but the high-frequency and long-term behaviour of
the gravitational wave power spectrum are both incorrect.

In Section~\ref{sec:production} we review the production of
gravitational waves by bubble collisions in the formalism of
Ref.~\cite{WeinbergBook}, and the approximations involved. Next, in
Section~\ref{sec:envelope} we give details of our numerical evaluation
of the envelope approximation, and present the comparison of our
results to those obtained previously.

The techniques used in direct numerical lattice simulations of the
phase transition are reprised in Section~\ref{sec:direct}, and
comparisons with the envelope approximation are made. We discuss the
results in Section~\ref{sec:discussion}.

\section{Production of gravitational radiation}
\label{sec:production}

The quantity of interest is typically the fraction of energy emitted
as gravitational waves per decade,
\begin{equation}
\Omega_\text{GW} = \omega \frac{\mathrm{d}
  E_\text{GW}}{\mathrm{d}\omega} \frac{1}{E_\text{tot}} \equiv 
\frac{\mathrm{d} E_\text{GW}}{\mathrm{d} \ln \omega} \frac{1}{E_\text{tot}}
\end{equation}
where $E_\text{tot}$ is the total energy.

The gravitational wave power radiated in a direction
$\hat{\mathbf{k}}$ at a frequency $\omega$ per unit solid angle
$\Omega$ is~\cite{WeinbergBook}
\begin{equation}
\label{eq:weinberg}
\frac{\mathrm{d}E_{\text{GW}}}{\mathrm{d}\Omega \mathrm{d}\omega} = 2 G \omega^2
\lambda_{ij,lm}(\hat{\mathbf{k}})\tau_{ij}^*(\hat{\mathbf{k}},\omega)
\tau_{lm}(\hat{\mathbf{k}},\omega),
\end{equation}
where $\Lambda_{ij,lm}$ is the projection tensor
\begin{multline}
\Lambda_{ij,lm}(\hat{\mathbf{k}}) \equiv \delta_{il}\delta_{jm} -
2\hat{\mathbf{k}}_j \hat{\mathbf{k}}_m \delta_{ij} +
\frac{1}{2}\hat{\mathbf{k}}_i \hat{\mathbf{k}}_j \hat{\mathbf{k}}_l
\hat{\mathbf{k}}_m \\
- \frac{1}{2} \delta_{ij}\delta_{lm} +
\frac{1}{2}\delta_{ij}\hat{\mathbf{k}}_l\hat{\mathbf{k}}_m +
\frac{1}{2} \delta_{lm}\hat{\mathbf{k}}_i \hat{\mathbf{k}}_j;
\end{multline}
and $\tau_{ij}(\hat{\mathbf{k}},\omega)$ is the Fourier transformed
stress-energy tensor
\begin{equation}
\tau_{ij}(\hat{\mathbf{k}},\omega) = \frac{1}{2\pi} \int \mathrm{d}t \,
e^{i\omega t} \int \mathrm{d}^3 x \, e^{-i\omega \hat{\mathbf{k}}\cdot
  \mathbf{x}} \tau_{ij}(\mathbf{x},t).
\end{equation}
For a scalar field $\phi$, the source is given by
\begin{equation}
\tau^\phi_{ij} = \partial_i \phi \partial_j \phi
\end{equation}
while for a relativistic fluid with energy $\epsilon$, pressure $p$,
relativistic gamma factor $W$ and 3-velocity $V_i$, the source is
\begin{equation}
\tau^\mathrm{f}_{ij} = W^2 (\epsilon + p)V_i V_j
\end{equation}
(the pieces proportional to the metric in the full stress-energy tensor
are pure trace and hence do not source gravitational waves).

In addition to the linearised gravity approximation that yields the
above expressions, two further simplifications are usually employed
when computing the resulting gravitational wave
power~\cite{Kosowsky:1992vn}. First, the collided portions of the
bubbles are neglected; this is what is \textsl{most strictly}
described as the envelope approximation.  Second, the bubble walls are
treated as sufficiently thin that the oscillatory part of the integral
$e^{-i\omega \hat{\mathbf{k}}\cdot \mathbf{x}}$ is approximately
constant in the region where $\tau_{ij}$ is nonzero.

The combination of the above approximations (often collectively
referred to simply as the `envelope approximation') has also been
applied to thermal phase transitions, where the scalar field is
coupled to a number of degrees of freedom that form a
plasma~\cite{Kamionkowski:1993fg}. One can compute an efficiency
factor $\kappa^\text{f}$ for the conversion of vacuum energy into
plasma kinetic energy~\cite{Kamionkowski:1993fg,Espinosa:2010hh}, and
if one assumes that the shell of fluid is thin then the envelope
approximation can be adapted to this case~\cite{Huber:2008hg}.

The resulting simple prediction of a broken power law form for the
power spectrum from gravitational waves at a first-order phase
transition has been widely adopted in the literature (see for example
Refs.~\cite{Caprini:2010xv,Kozaczuk:2014kva,Dorsch:2014qpa,Schwaller:2015tja}):
an approximately $\omega^3$ dependence at low frequencies, and an
approximately $\omega^{-1}$ dependence at high frequencies. The break
occurs at a characteristic length scale, believed to be set by ratio
of the inverse nucleation rate to the Hubble rate
$(H_*/\beta)$. Heuristically, the $\omega^3$ dependence can be seen as
the absence of structure (`white noise') on longer length scales,
while it has been argued that the approximate $\omega^{-1}$ dependence
at short length scales can be attributed to the size distribution of
bubbles. In the following section we will show that both power laws
are intrinsic features of the set of approximations outlined above,
irrespective of whether the bubbles are nucleated simultaneously or
with a physically-motivated exponential rate.

The collision of a pair of scalar field bubbles was treated
numerically in Ref.~\cite{Kosowsky:1991ua}. Since then, there have not
been any further attempts to perform direct numerical simulations that
compare the envelope approximation with dynamical vacuum scalar fields
or field-fluid systems modelling thermal phase transitions. In
Refs.~\cite{Hindmarsh:2013xza,Hindmarsh:2015qta}, a transient
$\omega^{-1}$ power law was seen for the gravitational wave power
spectrum at early times, which was attributed to the scalar field
collisions.
This assertion shall be tested in the present work.

\section{Computations in the envelope approximation}
\label{sec:envelope}

The envelope approximation consists of approximating the stress-energy
of the bubble wall with an infinitesimally thin shell, yielding
\begin{align}
\label{eq:tijenv}
\tau_{ij}(\hat{\mathbf{k}},\omega) & = \kappa \rho_\text{vac} v_\mathrm{b}^3
C_{ij}(\hat{\mathbf{k}},\omega) \\
C_{ij}(\hat{\mathbf{k}},\omega) &= \frac{1}{6\pi} \sum_n \int
\mathrm{d}t \, e^{i\omega(t-\hat{\mathbf{k}}\cdot \mathbf{x}_n) (t-t_n^3)}
  A_{n,ij}(\mathbf{k},\omega) \\
\label{eq:aenv}
A_{n,ij} (\hat{\mathbf{k}},\omega) & = \int_{S_n} \mathrm{d}\Omega
\, e^{-i\omega v_b (t-t_n) \hat{\mathbf{k}}\cdot \hat{\mathbf{x}}}
\hat{\mathbf{x}}_i \hat{\mathbf{x}}_j.
\end{align}
If the source under study is a system of intersecting fluid shells,
then the efficiency factor $\kappa=\kappa^\mathrm{f}$ can be computed
according the the procedure in Ref.~\cite{Espinosa:2010hh}. For scalar
field bubble walls, an effective scalar field efficiency factor
$\kappa^\phi$ can be calculated from the energy density on the bubble
walls and the surface area of the bubbles.

We evaluate Eqs.~(\ref{eq:tijenv}-\ref{eq:aenv}) using the method
given in Ref.~\cite{Huber:2008hg}. By choosing a system of cylindrical
coordinates such that $\hat{\mathbf{k}}$ is aligned with the $z$-axis, the
projected stress-energy tensor becomes
\begin{multline}
\label{eq:lambdaalign}
\Lambda_{ij,lm} \tau^*_{ij} \tau_{lm} \\ = \frac{1}{2} (\tau^*_{xx} -
\tau^*_{yy})(\tau_{xx} - \tau_{yy}) + \tau_{xy}^* \tau_{xy} + \tau^*_{yx}\tau_{yx},
\end{multline}
and the integrations over $C_{ij}$ and $A_{ij}$ become much
simpler. We need simply compute
\begin{align}
C_{\pm}(\omega) &= \frac{1}{6\pi} \,
 \sum_{n}^{N_\mathrm{b}} \int \mathrm{d} t \,
 e^{i\omega (t-z_n)} (t-t_n)^3 A_{n,\pm} (\omega, t) \\
  A_{n,\pm}(\omega,t) & = \int_{-1}^{1} \mathrm{d}z \,
  e^{-i v_\mathrm{b} \omega(t-t_n) z} B_{n,\pm}(z,t)
\end{align}
where
\begin{align}
B_{n,+} (z,t) & = \frac{(1-z^2)}{2} \int_{S'_n} \mathrm{d}\phi \,
\cos(2\phi), \\
B_{n,-} (z,t) & = \frac{(1-z^2)}{2} \int_{S'_n} \mathrm{d}\phi \,
\sin(2\phi).
\end{align}
See the appendix of Ref.~\cite{Huber:2008hg} for more
details.
The summation over $n$ is a sum over
all $N_\mathrm{b}$ bubbles in the simulation volume, and the integration region
$S'_n$ is the area on the surface of the $n$th bubble that does not
intersect with \textsl{any} other bubble.

\begin{figure}[tb]
\begin{centering}
\includegraphics[scale=0.5]{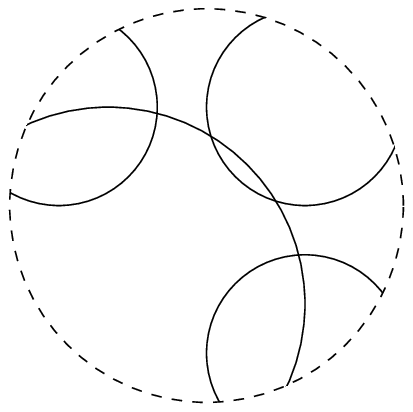}
\hfill
\includegraphics[scale=0.5]{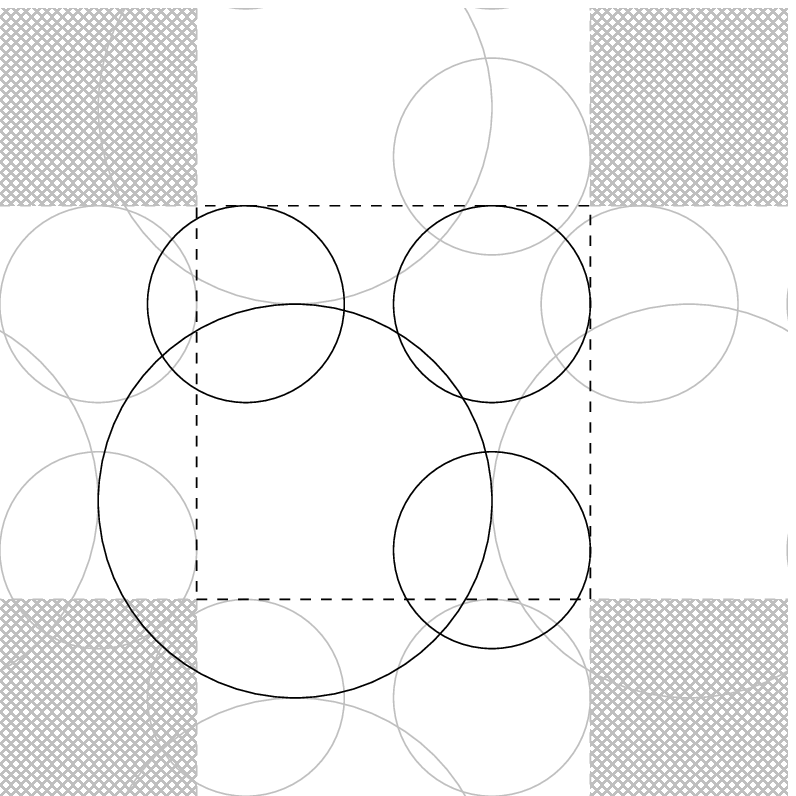}
\caption{\label{fig:geometry} Bubble geometries used in envelope
  approximation simulations. At left is the widely adopted spherical
  cutoff, where all gravitational wave power beyond a certain distance
  from the `central' bubble is ignored. At right is the `mirror'
  approach taken in the present work, where image bubbles are
  nucleated in neighbouring repeating unit cells; the aim of this is
  to closely model the periodic boundary conditions of lattice
  simulations. For a sufficiently large number of bubbles the two
  approaches are equivalent, corresponding to a system with `mirror'
  boundary conditions. 
}
\end{centering}
\end{figure}

To compute the envelope approximation for a fixed volume we not only
simulate the bubbles within the volume, but also follow the
development of bubbles in adjacent boxes (see
Fig.~\ref{fig:geometry}). We impose periodic boundary conditions on
our box, so we nucleate image bubbles in adjacent boxes.  These do not
contribute to the power but are included in the evaluation of the
uncollided bubble regions. This is in contrast to
Refs.~\cite{Huber:2008hg,Kosowsky:1992vn}, where a spherical volume is
used, but has the advantage that we can make truly direct comparisons
with lattice simulations. This means we must consider the interactions
of up to $27N_\mathrm{b}-1$ other bubbles when determining the
contribution of the $n$th bubble to the total power, although in
practice the number of bubbles in range is much smaller.

Our approach to computing the envelope approximation is therefore
about an order of magnitude more computationally intensive than
previous studies although the number of bubbles participating is
around the same. For comparisons with coupled field-fluid simulations,
this is not an issue as the dynamic range available there is a more
pressing constraint.

The fitted broken power law ansatz for the envelope approximation was
given in Eq.~(\ref{eq:huberfit}) -- a positive power law with
  index $a$ at low wavenumber and a negative power law with $b$
at high wavenumber. This also describes our own computations with the
envelope approximation and so curves given by fits to
Eq.~\ref{eq:huberfit} will be shown alongside our simulation results
in the following sections.

The theoretical expectation is that the low-frequency rising power law
has index $a = 3$, due to causality -- there is nothing in the system
on length scales larger than the largest bubble, so a cubic power law
is anticipated (two powers of $\omega$ from the radial integral, and
one additional power). We have confirmed this in our envelope
approximation simulations.

For the high-frequency power law, it is widely expected that $b
\approx 1$, either due to the size distribution of bubbles or
intrinsic effects. Unfortunately, we cannot reach high enough
frequencies $\omega$ to verify that $b$ is exactly unity.

However, we will show that these exponents are intrinsic to the
envelope approximation and do not depend on, for example, nucleation
rate. Later, we will also show that colliding scalar field bubble
walls give the same power laws.

\subsection{Testing the envelope approximation: geometry and nucleation rate}

It is standard to model the nucleation probability per unit volume and
time $P$ by
\begin{equation}
\label{eq:nucrate}
P = P_0 e^{\beta(t-t_0)},
\end{equation}
with $\beta$ computed, in principle, from the bounce
action~\cite{Coleman:1977py,Callan:1977pt,Linde:1981zj,Enqvist:1991xw}. We
instead take $\beta$ to be a constant numerical value throughout our
simulations in this section (in some cases we make it effectively
infinite: we nucleate bubbles simultaneously).

\begin{figure}[tb]
\begin{centering}
\includegraphics[width=0.45\textwidth,clip=true]{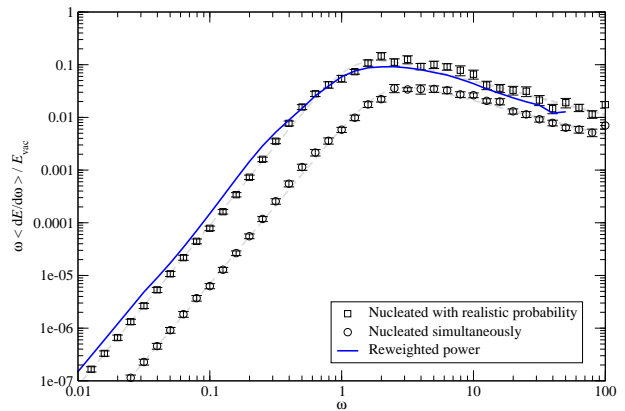}
\caption{\label{fig:comphuber} Comparison of scaled bubble collision
  power spectra, with $v_\mathrm{w}=1$. We show results from a
  simulation of 109 bubbles nucleated using the exponentially
  increasing nucleation rate (squares) and from one where the same
  number of bubbles are nucleated in the same positions simultaneously
  (circles). The parameters are such that comparison with Fig.~2 of
  Ref.~\cite{Huber:2008hg} is also possible, where the bubbles were
  nucleated at unequal times but with a spherical boundary to the
  simulation volume. As expected, there is no dependence on the form
  of the simulation volume. Furthermore, the unequal nucleation time
  case can be recovered from the equal nucleation time case by the
  reweighting outlined in the main text (solid blue
  curve).}
\end{centering}
\end{figure}

In this section, we consider the results of simulations with
$v_\text{w}=1$, 109 bubbles and $\beta=1$ or $\beta\to\infty$. For a
given bubble distribution, our results are the average of 32 uniformly
distributed random choices of the $z$-axis in
Eq.~(\ref{eq:lambdaalign}).

In Fig.~\ref{fig:comphuber}, we choose a single spatial distribution
of bubbles and nucleate them either over time with rate parametrised
by the `realistic' $\beta=1$, or simultaneously. For this bubble
distribution, a bootstrapped fit to Eq.~(\ref{eq:huberfit}) for
$\omega \in [0.01,100]$ yields $a=3.05 \pm 0.03$, $b=0.62 \pm 0.05$
for the `realistic' case. For the simultaneous case, we obtain $a=2.98
\pm 0.02$ and $b=0.65 \pm 0.04$. The power laws at high frequencies in
the envelope approximation are therefore not dependent on the size
distribution of bubbles at the end of the phase transition.

We have confirmed that these fitted power laws do not change
substantially when averaged over eight different bubble distributions.

We can do even more with the simultaneously nucleated
simulation. The resulting power spectrum can be rescaled to give the
gravitational wave power spectrum for a physical nucleation
rate. Given Eq.~(\ref{eq:weinberg}), we can write the rescaled
gravitational wave power as
\begin{multline}
\label{eq:reweight}
\omega \frac{\mathrm{d}E_{\text{GW}}}{\mathrm{d}\omega} = \omega
\sum_{n}^{N_\mathrm{b}} \frac{v_\mathrm{w}^3 (t_\text{end} -
  t_{0,n})^3}{\mathcal{V}} \\
\times 2 G \tilde{\omega}_n^2
\lambda_{ij,lm}(\hat{\mathbf{k}})\tau_{ij}^*\left(\hat{\mathbf{k}},
\tilde{\omega}_n\right)
\tau_{lm}\left(\hat{\mathbf{k}},\tilde{\omega}_n\right)
\end{multline}
with
\begin{equation}
\tilde{\omega}_n = \frac{v_\mathrm{w} (t_\text{end}
  - t_{0,n})}{(\mathcal{V}/N_\mathrm{b})^{1/3}}\omega,
\end{equation}
where $\mathcal{V}$ is the volume and the summation is over each
bubble, the $n$th bubble being nucleated at time $t_{0,n}$, the phase
transition ends at around $t_\text{end}$. The numerator is therefore
the approximate radius of the $n$th bubble, while the denominator
$(\mathcal{V}/N_\mathrm{b})^{1/3}$ is the average bubble radius in the
simulation where all bubbles were nucleated simultaneously. For
concreteness, we measure the remaining exposed surface area as a
function of time and take $t_\text{end}$ as the time when the surface area
of uncollided bubbles is less than 1\% of its peak value.

The result of applying this rescaling is also shown in
Fig.~\ref{fig:comphuber}, with good agreement. A similar rescaling
argument would presumably apply to the acoustic source in
Ref.~\cite{Hindmarsh:2015qta}. As the bubbles collide at different
times, fixing $t_\text{end}$ is an oversimplification, but it works
surprisingly well.

In summary, then, it is clear that the 
power laws seen in the envelope approximation are an intrinsic feature
of the calculation, rather than the distribution of bubbles. It is
also possible to reweight gravitational wave power spectra produced at
equal times to more realistic distributions.

\section{Direct simulations of the field-fluid system}
\label{sec:direct}

Having performed some tests of our new envelope approximation code
against the existing envelope approximation literature, we now wish to
make a comparison against the power spectra provided by lattice
simulations.

The equations and parameter choices we use have been discussed
extensively
elsewhere~\cite{Kajantie:1986hq,Ignatius:1993qn,Ignatius:1994fr,KurkiSuonio:1995pp,KurkiSuonio:1995vy,KurkiSuonio:1996rk,Hindmarsh:2013xza,Hindmarsh:2015qta},
so we present only a brief summary here. We are working with a coupled
system of a relativistic ideal fluid $U^\mu$ and scalar field $\phi$,
with energy-momentum tensor
\begin{equation}
\label{eq:tmunu}
T^{\mu\nu} = \partial^\mu \phi \partial^\nu \phi - {\textstyle \frac{1}{2}} g^{\mu\nu} (\partial\phi)^2
+ \left[\epsilon + p \right] U^\mu U^\nu + g^{\mu\nu} p,
\end{equation}
where the metric is $g^{\mu\nu} = \mathrm{diag}(-1,1,1,1)$. The system
has an effective potential
\begin{equation}
\label{eq:effpot}
V(\phi, T) = \frac{1}{2} \gamma (T^2-T_0^2) \phi^2 - \frac{1}{3} \alpha T \phi^3 + \frac{1}{4}\lambda\phi^4,
\end{equation}
with parameters given in Table~\ref{tab:parameters}.
The equation of state is
\begin{align}
\epsilon(T,\phi) &= 3 a T^4 + V(\phi,T) - T\frac{\partial V}{\partial T},\\
  p(T,\phi) &= a T^4 - V(\phi,T),
\end{align}
$a = (\pi^2/90)g_*$; we take $g_*=34.25$ for consistency with previous
papers.  From the energy-momentum tensor we can derive equations of
motion. The system is decomposed into our choice of `field' and
`fluid' parts,
\begin{align}
  [\partial_\mu T^{\mu\nu}]_{\rm field} &=
    (\partial_\mu\partial^\mu\phi) \partial^\nu\phi -
    \frac{\partial V}{\partial\phi} \partial^\nu \phi  \nonumber \\ 
   &  \qquad = \eta U^\mu \partial_\mu\phi \partial^\nu \phi,\\
    [\partial_\mu T^{\mu\nu}]_{\rm fluid} &=
   \partial_\mu[(\epsilon+p)U^\mu U^\nu] - \partial^\nu p +
    \frac{\partial V}{\partial \phi} \partial^\nu \phi  \nonumber \\
   & \qquad =  -\eta U^\mu \partial_\mu\phi \partial^\nu \phi.
\end{align}
The parameter $\eta$ sets the scale of the friction and, hence, the
wall velocity.

We simulate the coupled field-fluid system using parameters familiar
from Refs.~\cite{Hindmarsh:2013xza,Hindmarsh:2015qta}, summarised in
Table~\ref{tab:parameters} (again, for consistency with previous work,
we use units where $G=T_\mathrm{c}=1$). The `weak' and `weak scaled'
parameters give a phase transition strength $\alpha_{T_\mathrm{N}}
\approx 0.01$, while the `intermediate' parameters give
$\alpha_{T_\mathrm{N}} \approx 0.1$.

\begin{table}
\begin{tabular}{l | c | c | c }
 & Weak & Weak (scaled)  & Intermediate \\
\hline
$T_0/T_\mathrm{c}$ & $1/\sqrt{2}$ & $1/\sqrt{2}$ & $1/\sqrt{2}$ \\
$\gamma$ & $1/18$ & $4/18$ & $2/18$ \\
$\alpha$ & $\sqrt{10}/72$ & $\sqrt{10}/9$ & $\sqrt{10}/72$ \\
$\lambda$ & $10/648$ & $160/648$ & $5/648$ \\
\end{tabular}

\caption{\label{tab:parameters} Potential parameters used for this
  paper. These are the same as for
  Refs.~\cite{Hindmarsh:2013xza,Hindmarsh:2015qta}. The `weak scaled'
  parameters give the same phase transition strength as the `weak'
  parameters, but the bubble wall width is halved (see Table~\ref{tab:efficiencies}).}
\end{table}

It was hoped that `scaled' forms of the `intermediate' parameters could
also be used, to improve the dynamic range of the simulations, but in tests
using a spherically symmetric code it was found that the resulting
fluid shock at $v_\mathrm{w} = 0.44$ cannot be resolved well with
  our current simulation code and available resources.

From the potential~(\ref{eq:effpot}), we can compute the surface
tension $\sigma$
\begin{equation}
\label{eq:sigma}
  \sigma = \frac{2\sqrt{2}}{81}\frac{\alpha^3}{\lambda^{5/2}} T_\mathrm{c}^3
\end{equation}
and the correlation length in the broken phase $\ell$
\begin{equation}
  \ell^2 = \frac{9\lambda}{2\alpha^2} \frac{1}{T_\mathrm{c}^{2}}.
\end{equation}
The latent heat at the critical temperature is
\begin{equation}
\mathcal{L} = \frac{\alpha^2 \gamma}{\lambda^2} T_0^2 T_\mathrm{c}^2
\end{equation}
and the ratio of latent heat to radiation $\alpha_{T}$ can then be
written as
\begin{equation}
\label{eq:alphat}
\alpha_{T} = \frac{\mathcal{L}}{3 a T^4}.
\end{equation}
These derived quantities, along with other relevant quantities for the
simulations, are shown in Table~\ref{tab:efficiencies}.

\subsection{Gravitational waves from colliding scalar field bubbles}
In Ref.~\cite{Hindmarsh:2015qta} it was conjectured that the
$\omega^{-1}$ power law seen above the peak in the gravitational wave
power spectrum was the same as that produced by the envelope
approximation. In this section we shall test that hypothesis.

To make the comparison as exact as possible we compare bubbles
nucleated in exactly the same positions for both the field-fluid model
and the envelope approximation, although we note that even for
$N_\mathrm{b}=37$ in Ref.~\cite{Hindmarsh:2015qta} there was no
noticeable difference between power spectra when bubbles were
nucleated in different positions. Again, since the envelope
approximation calculation outlined above only gives the power radiated
in the specified $\hat{\mathbf{z}}$-direction, we repeat the envelope
approximation simulation for 32 randomly selected directions uniformly
distributed on the surface of the sphere. As we are directly
  comparing lattice simulations and envelope approximation
  calculations for the same configuration, random errors quoted in
  this section are those arising from this sampling.

\begin{figure}[tb]
\begin{centering}
\includegraphics[width=0.45\textwidth,clip=true]{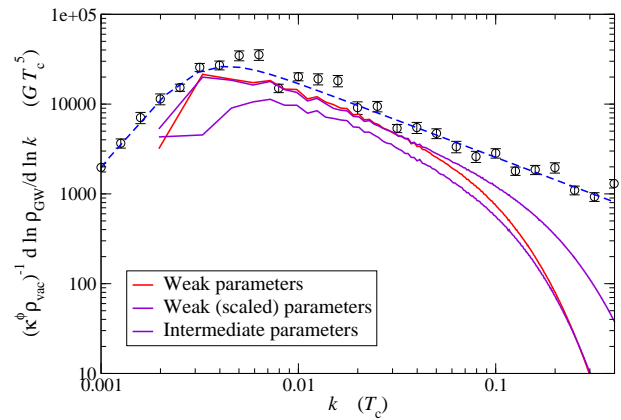}
\caption{\label{fig:unscaled} Gravitational waves from colliding
  scalar field bubble walls. Comparison of power spectra computed by
  the envelope approximation (points with error bars; blue dashed line
  fit) and by lattice simulations with source $\tau_{ij}^\phi$ and
  `weak' (red curve), `weak scaled' (purple curve), and `intermediate'
  (green curve) phase transition parameters given in the main
  text. The resulting gravitational wave power is scaled by the scalar
  field gradient energy density $\kappa^\phi \rho^\text{vac}$, meaning
  all three lattice simulations are directly comparable to one
  envelope computation. The envelope computations were made with
  the same bubble positions, asymptotic wall velocity
  ($v_\mathrm{w}=0.44$), and scalar gradient energy as develops during
  the lattice simulation.}
\end{centering}
\end{figure}

In Fig.~\ref{fig:unscaled}, we rescale the gravitational wave power
spectra for the `weak', `weak scaled' and `intermediate' phase
transition parameters by the respective scalar field energy
densities. This shows that the `weak' and `weak scaled' cases give
broadly similar power spectra, lying systematically about $1\sigma$
below most points in the envelope approximation. Agreement with the
`intermediate' parameters is not as good, perhaps because the increased
surface tension deforms the bubbles as they collide.

\begin{figure}[tb]
\begin{centering}
\includegraphics[width=0.45\textwidth,clip=true]{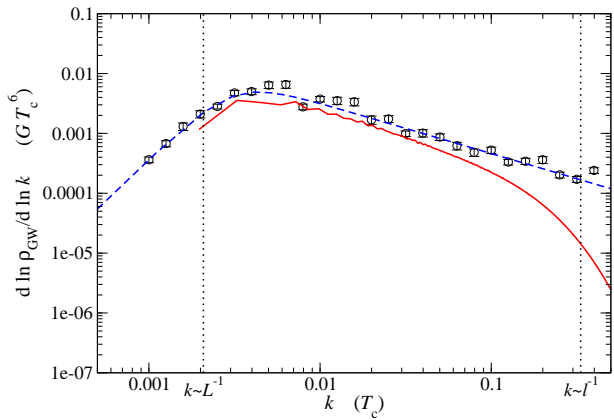}
\caption{\label{fig:complatenvelope} As for Fig.~\ref{fig:unscaled},
  but without the scaling by $\kappa^\phi \rho^\text{vac}$, so
  comparison with only one lattice simulation is possible, in this
  case the `weak scaled' simulation. The box size $L$ and approximate
  wall width $\ell$ are shown by vertical dashed lines.}
\end{centering}
\end{figure}

\begin{figure}[tb]
\begin{centering}
\includegraphics[width=0.45\textwidth,clip=true]{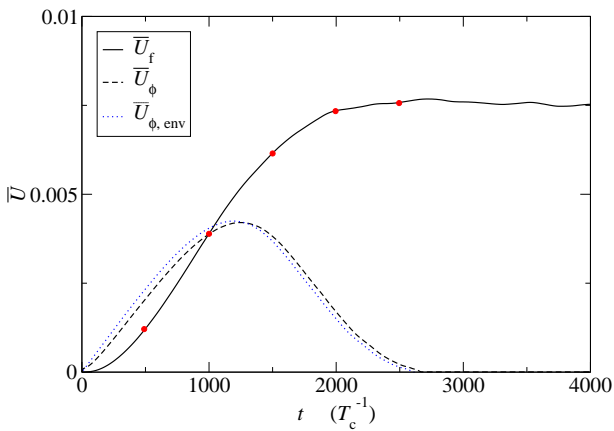}
\caption{\label{fig:ubars} Dimensionless estimates of the field and
  fluid energy density ($\overline{U}_f$ and $\overline{U}_\phi$) for
  the `weak scaled' simulation (see Fig.~\ref{fig:complatenvelope}),
  along with inferred field energy density from the corresponding
  envelope approximation calculation
  ($\overline{U}_{\phi,~\text{env}}$). The red circles indicate
    the times at which the gravitational wave power spectrum sourced
    by the fluid is shown in Fig.~\ref{fig:complatfluid}.}
\end{centering}
\end{figure}

In Fig.~\ref{fig:complatenvelope}, we instead scale the envelope
approximation and compare with the actual gravitational wave power
spectrum for the `weak scaled' parameters. Also shown are the
broken-phase correlation length $\ell$ and the simulation box size
$L$, to give an indication of the limitations imposed by dynamic
range. For this case, in Fig.~\ref{fig:ubars} we also plot the
dimensionless scalar field and fluid kinetic energy quantities:
\begin{align}
\label{eq:ubarf} 
\overline{U}_\mathrm{f} &
=\sqrt{\frac{1}{\mathcal{V}(\overline{\epsilon} +
    \overline{p})}\int d^3 x \, \tau_{ii}^\mathrm{f}}, \\
 \overline{U}_\phi & = \sqrt{\frac{1}{\mathcal{V}(\overline{\epsilon} +
    \overline{p})}\int d^3 x \, \tau_{ii}^\phi} .
\end{align}
These are compared to the equivalent quantity
$\overline{U}_{\phi,\text{env}}$ computed during a run of the envelope
approximation based on the uncollided surface area and the scalar
field gradient energy. There is very good agreement between
$\overline{U}_\phi$ and $\overline{U}_{\phi,\text{env}}$: the
acceleration of the bubbles in the full simulation is seemingly of
little importance to the scalar field source.

For the phase transition strengths studied here there is no evidence
that the presence of fluid in front of the wall affects the power
spectrum. The gravitational wave power spectrum sourced by the scalar
field, shown in
Figs.~\ref{fig:unscaled}~and~\ref{fig:complatenvelope}, does not
change when the fluid part of the simulation is not evolved
dynamically but the friction term is retained so that the wall
velocity is unchanged.

\subsection{Gravitational waves from colliding fluid shells}

\begin{figure}[tb]
\begin{centering}
\includegraphics[width=0.45\textwidth,clip=true]{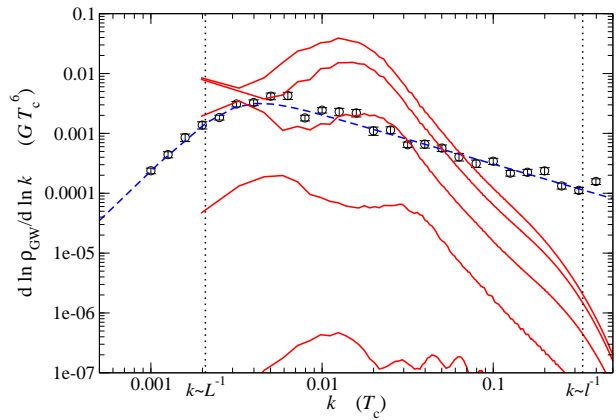}
\caption{\label{fig:complatfluid} Gravitational waves from colliding
  fluid shells for a deflagration. The envelope approximation result
  (blue dashed fit, black circle measurements) is scaled with the
  fluid energy density $\kappa^\mathrm{f} \rho^\text{vac}$, with the
  peak value agreeing well with Eq.~\ref{eq:peakfreq}.  The red curves
  show the gravitational wave power spectrum from the `weak scaled'
  lattice simulation sourced by $\tau_{ij}^\text{f}$ at intervals of
  $500/T_\mathrm{c}$. 
}
\end{centering}
\end{figure}

We now turn our attention to the form of the fluid power spectrum
immediately after the bubbles have collided, and hypothesise that the
gravitational wave power produced by the fluid up to this point might
still be computed with the envelope approximation. The later
contribution due to sound waves must then be computed separately.

The computation proceeds as before, except that in
Eq.~(\ref{eq:tijenv}) the energy density is scaled by the fraction
$\kappa^\mathrm{f}$ of vacuum energy that gets turned into fluid
kinetic energy as the bubbles grow. While in the scalar field case in
the previous section, the energy density $\rho$ was estimated based on
the surface of each bubble at collision, here we can directly measure
the total fluid kinetic energy at the end of the transition and use it
to compute an efficiency (see Table~\ref{tab:efficiencies}); this can
also be computed analytically~\cite{Espinosa:2010hh}.

\begin{table}
\begin{tabular}{l | c | c | c }
 & Weak & Weak (scaled)  & Intermediate \\
\hline
$\sigma/T_\mathrm{c}^3$ & $1/10$ & $1/20$ & $4\sqrt{2}/10$ \\
$\ell \, T_\mathrm{c}$ & $6$ & $3$ & $6 \sqrt{2}$ \\ 
$\mathcal{L}/T_\mathrm{c}^4$ & $9/40$ & $9/40$ & $9/5$ \\
$\alpha_{T_\mathrm{N}}$ & $0.010$ & $0.010$ & $0.084$ \\
\hline
$\mathcal{V} \, T_\mathrm{c}^3$ & $4800^3$ & $4800^3$ & $4800^3$ \\
$N_\mathrm{b}$ & $125$ & $125$ & $125$ \\
$R_*$ & $960$ & $960$ & $960$ \\
$\mathcal{S} \, T_\mathrm{c}^2$ & $6.98 \times 10^8$ & $6.98 \times 10^8$ & $6.98 \times
10^8$ \\
\hline
$\eta/T_\mathrm{c}$ & $0.2$ & $0.4$ & $0.4$ \\
$v_\mathrm{w}$ & $0.44$ & $0.44$ & $0.44$ \\
\hline
$\Sigma$ & $0.136$  & $0.068$ & $0.776$ \\
$\kappa^\phi \rho^\text{vac}/T_\mathrm{c}$ & $0.000859$ & $0.000430$ & $0.00490$ \\
$\kappa^\mathrm{f}$ & $0.0348$ & $0.0348$ & $0.195$ \\
$\kappa^\mathrm{f}  \rho^\text{vac}/T_\mathrm{c}$ & $0.000348$ & $0.000348$ & $0.0164$  \\
\end{tabular}

\caption{\label{tab:efficiencies} Table of various derived quantities
  and simulation parameters. The surface tension $\sigma$, broken
  phase correlation length $\ell$ (approximately the bubble wall
  thickness) and phase transition strength $\alpha_{T_\mathrm{N}}$ are
  computed from the parameters in Table~\ref{tab:parameters} by means
  of Eqs.~(\ref{eq:sigma}-\ref{eq:alphat}). The simulation volume
  $\mathcal{V}$ and number of bubbles $N_\mathrm{b}$ are set for each
  simulation, yielding the typical bubble radius $R_* = (\mathcal{V}/N_b)^{1/3}$. The total collided `surface area' $\mathcal{S}$
  can be computed given the Voronoi partition of the nucleated bubble
  locations. The friction parameter $\eta$ determines the wall
  velocity $v_\mathrm{w}$, chosen in each case to give a fast
  deflagration (such that $v_\mathrm{w} \approx 0.44$ would be
  achieved if the bubbles could expand unimpeded). The scalar field
  gradient energy per unit area $\Sigma$ is approximately proportional
  to $\sigma$. The scalar field energy density is then $\kappa^\phi
  \rho^\text{vac} = \Sigma \mathcal{S}/\mathcal{V}$, since the energy
  in the scalar field source scales only with the radius of the
  bubbles, whereas the fluid source scales with the volume and hence
  $\kappa^\mathrm{f} \rho^\text{vac}$ can be computed analytically for
  the general case~\cite{Espinosa:2010hh}.  }
\end{table}

In Fig.~\ref{fig:ubars} the dimensionless measure of the fluid kinetic
energy density $\overline{U}_\mathrm{f}$ is shown, and as expected it
remains approximately constant after the phase transition
completes. Consequently, the acoustic waves present in the fluid
continue to source gravitational waves, and in
Fig.~\ref{fig:complatfluid} the amplitude of the gravitational waves
from the lattice simulation -- shown in red -- continues to grow after
the phase transition has completed. The envelope approximation result,
appropriately scaled by $\kappa^{f}\rho^\text{vac}$, is also
shown.

By comparing the fluid kinetic energy at the indicated time intervals
of $500/T_\mathrm{c}$ in Fig.~\ref{fig:ubars} with the succession of
red curves in Fig.~\ref{fig:complatfluid} -- the fourth point and
curve in particular -- one can see that the envelope approximation
gets the amplitude of the gravitational wave power produced by
colliding fluid shells correct to within an order of magnitude at the
time the fluid kinetic energy has reached its final value. The form of
the power spectrum is, however, different; the peak is offset; and the
total power continues to grow after this time, sourced by acoustic
waves set up in the plasma after collision. 

To be specific, up to the time at which the bubbles collide,
  there are two peaks -- one closer to the infrared at $k_\text{env} \sim 1/R_*
  $, and another at higher $k$ around the fluid shell thickness $k_\text{shell}
  \sim v_\mathrm{w}/(R_* | v_\mathrm{w} - c_s|)$. In this case, $k_\text{shell
   } \approx 3 k_\text{env}$. The $k_\text{shell}$ peak
  continues to grow, sourced by acoustic waves in the fluid. Note also
  that the high-$k$ power law associated with this peak is steeper
  (approximately $k^{-3}$) than the envelope approximation. This peak
  continues to grow until extinguished by expansion on a timescale
  $1/H_*$~\cite{Hindmarsh:2015qta}.

As a final note, in the $v_\mathrm{w} \approx 0.44$, `weak parameter'
simulations discussed extensively here, the amplitude of gravitational
waves sourced by the scalar field (shown in
Fig.~\ref{fig:complatenvelope}) and by the fluid (show in
Fig.~\ref{fig:complatfluid}) are comparable, at least as the phase
transition is ending. However, in a realistic scenario, the scalar
field bubble walls would be much thinner than the fluid shell, and in
any case, acoustic waves (and possibly turbulence) play a more
important role, at least for thermal phase transitions.

\section{Discussion}
\label{sec:discussion}

We have revisited previous work employing the envelope approximation
to compute the gravitational wave power spectrum from colliding
bubbles. The observed power laws -- which are consistent with
what was seen in Ref.~\cite{Huber:2008hg} -- do not depend on the
nucleation rate, and it is possible to reweight from simulations with
simultaneous nucleation to more physical scenarios. We have no evidence
that the hierarchy of bubble sizes affects the power law above the
peak -- the power laws are an intrinsic feature of the envelope
approximation.

In addition, we compared the envelope approximation to large-scale
lattice simulations of a thermal phase transition, where a scalar
field expands in a plasma of light particles. The envelope
approximation is a good model for the gravitational waves produced by
the scalar field, although it seems to perform less well at higher
surface tensions.

For gravitational waves sourced by the plasma, the envelope
approximation gets the peak amplitude approximately correct, but the
form of the power spectrum is incorrect. Furthermore, the
subsequent acoustic and turbulent behaviour cannot be modelled at all.

This paper focused on fairly fast ($v_\mathrm{w} = 0.44$) subsonic
deflagrations. In such cases the fluid shells are very thick, about a
third the radius of the bubble itself. Future work will consider the
possibility that the initial transient collision of very thin fluid
shells -- such as in the fine-tuned Jouguet case where $v_\mathrm{w}
\approx c_\mathrm{s}$ -- might be better described by the envelope
approximation. At late times, though, the dominant sources will still
be sound waves and turbulence.

It is expected that, for a viable first order electroweak-scale phase
transition, the sources for which the envelope approximation would be
valid -- the scalar field bubble collisions, and possibly the initial
fluid shell collisions -- are subdominant. However, this is not
necessarily the case, depending on $\beta/H_*$, or if the bubble wall
runs away. Our results are valuable in any case because they represent
the first comparison of the envelope approximation with alternative
methods of modelling gravitational waves. They also represent an
external test of the `gravity sector' of the simulation code in
Refs.~\cite{Hindmarsh:2013xza,Hindmarsh:2015qta}.

We used novel boundary conditions for our envelope approximation
calculation. When the `spherical cutoff' approach is used, the number
of pairs of bubble interactions that must be checked grows only as
$N_\mathrm{b}^2$, whereas in our hypercubic case, it grows as $27
N_\mathrm{b}^2$.  Therefore, if we had adopted the standard
technique, about five times as many bubbles could be simulated
for the same amount of computing time.

There are important consequences for modelling the gravitational
  wave production from first-order phase transitions: the true peak
  may be shifted to higher $k$ by up to an order of magnitude,
  although the amplitude will be higher; and the power laws associated
  with the peak may well be steeper. These effects mean care must be
  taken when discussing the prospects for detection at future
  detectors such as eLISA~\cite{Caprini:2015zlo}.

We conclude by reaffirming the utility of the envelope approximation
for modelling the immediate aftermath of a thermal phase transition,
or for situations where the fluid does not contribute (such as vacuum
bubbles). For the majority of cases -- where the fluid source is
dominant -- an analytic or at least semi-analytic method is still
  lacking, and direct numerical simulation is still necessary.

\begin{acknowledgments}
Our simulations made use of `gorina1' at the University of Stavanger
as well as the Abel cluster, a Notur facility. We acknowledge PRACE
for awarding us access to resource HAZEL HEN based in Germany at the
High Performance Computing Center Stuttgart (HLRS). We acknowledge
useful discussions with Mark Hindmarsh, Stephan Huber, Kari
Rummukainen and Anders Tranberg. Our work was supported by the People
Programme (Marie Sk{\l}odowska-Curie actions) of the European Union
Seventh Framework Programme (FP7/2007-2013) under grant agreement
number PIEF-GA-2013-629425.
\end{acknowledgments}

\renewcommand{\emph}[1]{\textsl{ #1 } } 

\bibliography{comparison}

\end{document}